\documentclass[debug,overfull,figures,preprint]{epl}

\usepackage{amsmath}
\usepackage{amssymb}
\usepackage{mathrsfs}
\usepackage{graphics}
\usepackage{graphicx}
\usepackage{epsf}

\newcommand{\tr}{\mathsf{t}}

\title{Stochastic approach to DNA breathing dynamics}
\author{Suman Kumar Banik \and Tobias Ambj\"ornsson \and 
        Ralf Metzler\thanks{E-mail: \email{metz@nordita.dk}}}
\shortauthor{S. K. Banik \etal}
\institute{NORDITA, Nordic Institute for Theoretical Physics -
           Blegdamsvej 17, DK-2100 Copenhagen {\O}, Denmark}


\begin{document}

\maketitle

\begin{abstract}
We propose a stochastic Gillespie scheme to describe the temporal fluctuations
of local denaturation zones in double-stranded DNA as a single molecule time
series. It is demonstrated that the model recovers the equilibrium properties.
We also study measurable dynamical quantities such as the bubble size
autocorrelation function. This efficient computational approach will be useful
to analyse in detail recent single molecule experiments on clamped homopolymer
breathing domains, to probe the parameter values of the underlying
Poland-Scheraga model, as well as to design experimental conditions for
similar setups.
\end{abstract}

\section{Introduction}

\begin{figure}
\twoimages[scale=0.22,angle=0]{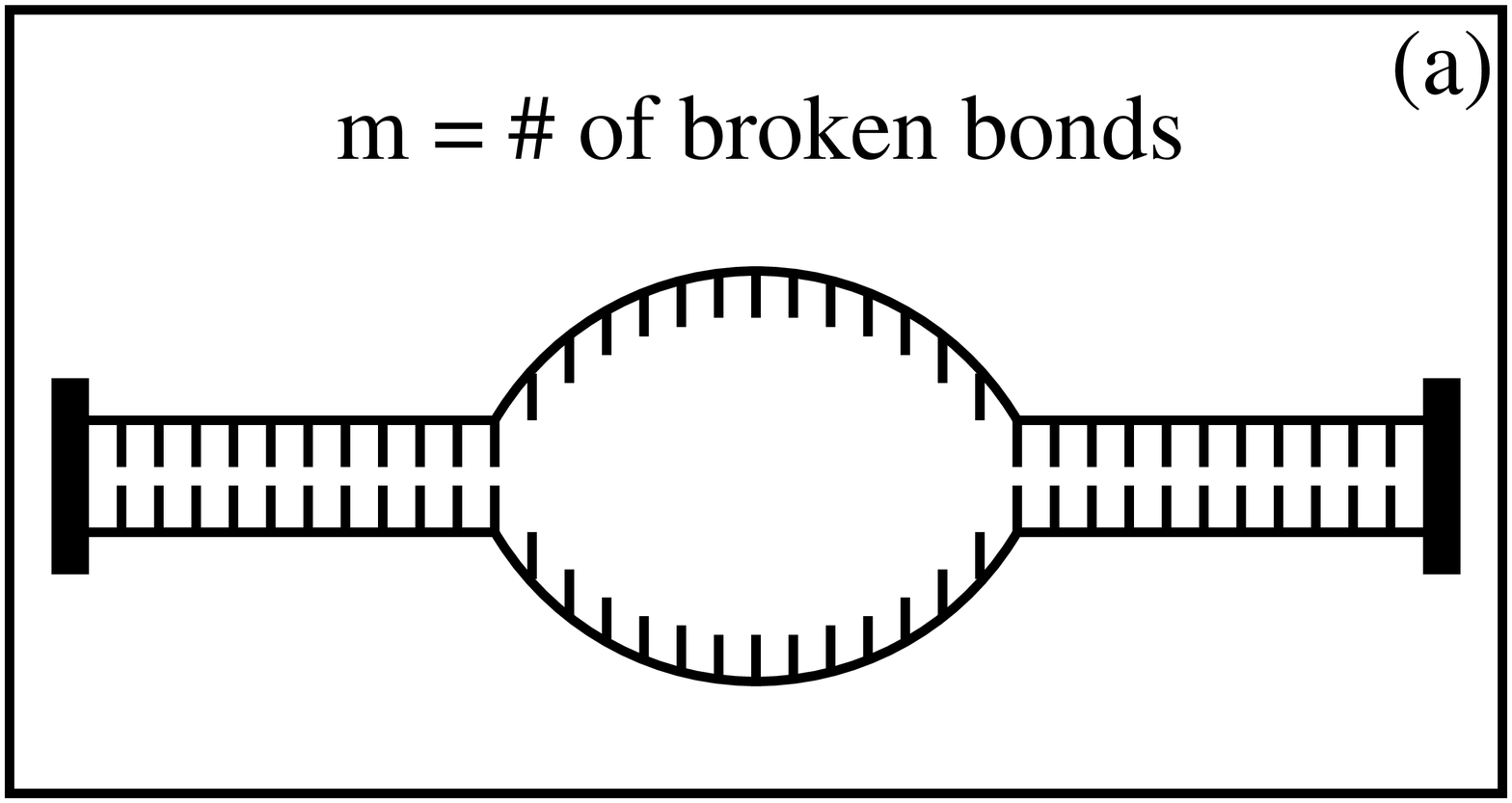}{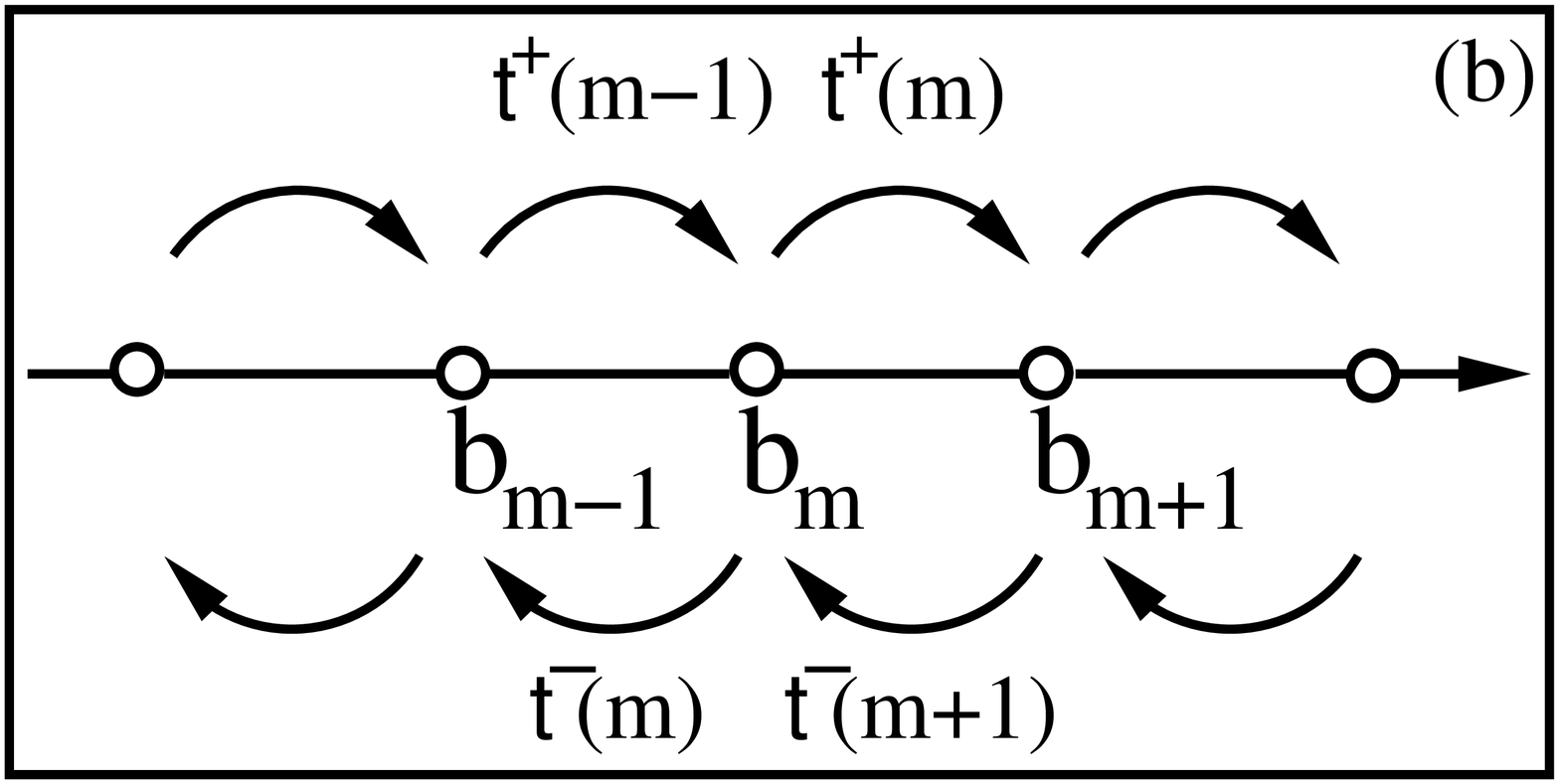}
\caption{(a) A stretch of dsDNA that is clamped at both ends, with an ssDNA
bubble consisting of $m$ broken bps. (b) Schematic representation of
the jump process underlying bubble-breathing.}
\label{fig1}
\end{figure}

Under a large range of salt conditions and temperatures, the double-helix
is the thermodynamically stable configuration of DNA \cite{kornberg,poland}.
This stability is effected by the Watson-Crick H-bonding of base-pairs (bps),
and the stronger base-stacking of neighbouring, planar aromatic bps, that 
by hydrophobic interactions stabilize the helical structure
\cite{watsoncrick,delcourt}.
At the same time, double-stranded DNA (dsDNA) is distinguished by the
ease with which locally the molecule can open up (and later rejoin), to
produce flexible \emph{bubbles} of single-stranded DNA (ssDNA), see
figure \ref{fig1}a \cite{kornberg,poland}. The formation of DNA-bubbles,
despite the rather large enthalpy necessary to break the base-stacking,
is made possible by the entropy gain, due to which the free energy for
breaking additional bps is of the order of $k_BT$, after overcoming
a bubble \emph{initiation barrier} $\sigma_0\simeq 10^{-5}\ldots
10^{-3}$ \cite{blake}.
At room or physiological temperature, bubbles of 20 to 30 broken
bps are created \cite{poland}. DNA-bubbles preferentially
form in regions rich in the weaker AT bps \cite{blake}, and they
are related to physiological processes such as transcription initiation
\cite{kornberg1}.

Traditionally measured in bulk by UV-absorption at elevated temperatures
\cite{wartell}, it is now possible to probe the time series of the size
fluctuations of a single DNA-bubble (\emph{DNA-breathing}) by fluorescence
correlation techniques in short, designed DNA-stretches \cite{altan}.
These DNA-constructs contain a well-defined (AT)$_M$ breathing domain, that
is labelled by a fluoro\-pho\-re-quencher pair. Such a model system allows for
a precise quantitative analysis, from which the parameters of the underlying
theoretical model (usually the Poland-Scheraga model) can be determined.
This fluorescence technique is being developed further to measure longer
bubble domains, also at higher temperatures. It is therefore of interest to
provide a theoretical model to understand DNA-breathing in a
homopolymer domain quantitatively. One possible approach is based on
the master equation for the probability distribution to find a certain
bubble size at time $t$, from which quantities such as the mean bubble
size or the bubble autocorrelation function can be derived
\cite{ambme,hame,ambme1}. Despite its mathematically appealing formulation,
the master equation needs to be solved numerically by inverting the
transfer matrix \cite{ambme,ambme1}. Moreover, it produces
ensemble-averaged information. Given the access to single molecule data,
it is of relevance to obtain a model for the fully stochastic time evolution
of a single DNA-bubble, providing a description for pre-noise-averaged
quantities such as the step-wise (un)zipping. With this scope, we
here introduce a stochastic simulation scheme for the (un)zipping dynamics.
We use the Gillespie algorithm to update the state of the system by determining
(i) the random time between individual (un)zipping events, and (ii) which
reaction direction (zipping, $\leftarrow$, or unzipping, $\rightarrow$) will
occur. We corroborate that the model recovers the equilibrium properties, and
study the single bubble time evolution. The proposed scheme is efficient
computationally, easy to implement, and amenable to generalization.

\section{The Model}

Motivated by the \emph{in vitro\/} study of reference \cite{altan}, we
consider a dsDNA segment with $M$ bps that
is clamped at both ends (figure \ref{fig1}a).
A bubble state of $m$ broken bps is defined by the occupation
numbers $b_m=1$ and $b_{m'}=0$ ($m'\neq m$). The stochastic simulation
then corresponds to the nearest-neighbour jump process (figure \ref{fig1}b)
\begin{equation}
\label{eq1}
b_0\rightleftarrows b_1\rightleftarrows\ldots
\rightleftarrows b_m\rightleftarrows\ldots
\rightleftarrows b_{M-1}\rightleftarrows b_M,
\end{equation}
with reflecting boundary conditions at $b_0$ and $b_M$. Each jump away
from state $b_m$ occurs after a random time $\tau$, and in random direction
to either $b_{m-1}$ or $b_{m+1}$, governed by the reaction probability
density function (PDF)\footnote{The original expression
$P(\tau,\mu)=b_m\tr^{\mu}(m)\exp\left(-\tau\sum_{m,\mu}b_m\tr^{\mu}(m)
\right)$, that is relevant for consideration of multi-bubble states,
simplifies here due to the particular choice of the
$b_m$.} \cite{gillespie}
\begin{equation}
\label{rpdf}
P(\tau,\mu)=\tr^{\mu}(m)e^{-\left(\tr^+(m)+\tr^-(m)\right)\tau},
\end{equation}
where $\mu\in\{+,-\}$ denotes the unzipping ($+$) or zipping ($-$) of a
bp, and the
jump rates $\tr^{\pm}(m)$ are defined below. From the joint PDF (\ref{rpdf}),
the waiting time PDF that a jump away from $b_m$ occurs is given by $\psi(\tau)
=\sum_{\mu}P(\tau,\mu)$, i.e., it is Poissonian. The probability that the
bubble size does not change in the time interval $[0,t]$ is given by
$\phi(t)=1-\int_0^t\psi(\tau)d\tau$.

The rates $\tr^{\pm}(m)$ are based on the statistical weight for a
homopolymer bubble \cite{poland,richard}
\begin{equation}
\label{eq2}
\mathscr{Z}^{\Bumpeq}(m)=\sigma_0 u^m(1+m)^{-c}
\end{equation}
according to the Poland-Scheraga model, with $\mathscr{Z}^{\Bumpeq}(0)=1$.
Here, we introduced the loop initiation factor $\sigma_0$ for breaking the
initially unperturbed dsDNA; the statistical weight $u=\exp(-E/k_BT)$
with the free energy $E=E(T)$ associated with breaking an additional
bp. For the (AT) domain (with melting temperature $67^{\circ}$C), $u=0.6$
corresponds to physiological temperature, and $u=0.9$ to $\approx 61^{\circ}$C.
We note that in single molecule stretching experiments the effective
temperature can be changed by applying an external torque ${\mathfrak T}$,
such that $u\to u\exp(\theta_0\mathfrak{T}/k_BT)$, where $\theta_0=2\pi/
10.35$ is the twist angle per bp of the dsDNA \cite{cocco,pant}.
Equation (\ref{eq2}) also introduces the loop closure factor $(1+m)^{-c}$
assigned to the entropy loss for creating a polymer loop of size $m$; the
offset by 1 is due to persistence length corrections \cite{blake,fixman}.
For the loop closure exponent in the rather short DNA-segment we have in
mind here, we choose $c=1.76$ \cite{fisher,blake,comment}, although higher
values have been suggested \cite{richard}. Requiring that the system
eventually reaches equilibrium, the rates $\tr^{\pm}(m)$ have to fulfil
the detailed balance condition $\tr^+(m-1)\mathscr{Z}^{\Bumpeq}(m-1)=
\tr^-(m)\mathscr{Z}^{\Bumpeq}(m)$. This condition does not fully specify
the rates, and we choose the specific form \cite{ambme,ambme1}
\begin{equation}
\label{eq3}
\tr^+(0)=2^{-c}k\sigma_0u, \,\, \tr^+(m)=k\frac{\mathscr{Z}^{\Bumpeq}(m+1)}{
\mathscr{Z}^{\Bumpeq}(m)}=ku\left(\frac{1+m}{2+m}\right)^c, \,\, \tr^-(m)=k,
\end{equation}
completed by $\tr^-(0)=\tr^+(M)=0$. This choice assumes that the
unzipping of a bp is limited by the Boltzmann factor $u$ ($\sigma_0u$ for
bubble initiation), whereas bp-zipping 
occurs with constant rate $k$, specified by the sterical formation of
a bp (to be determined from microscopic models).

\section{Results and discussions}

\begin{figure}
\onefigure[scale=0.47]{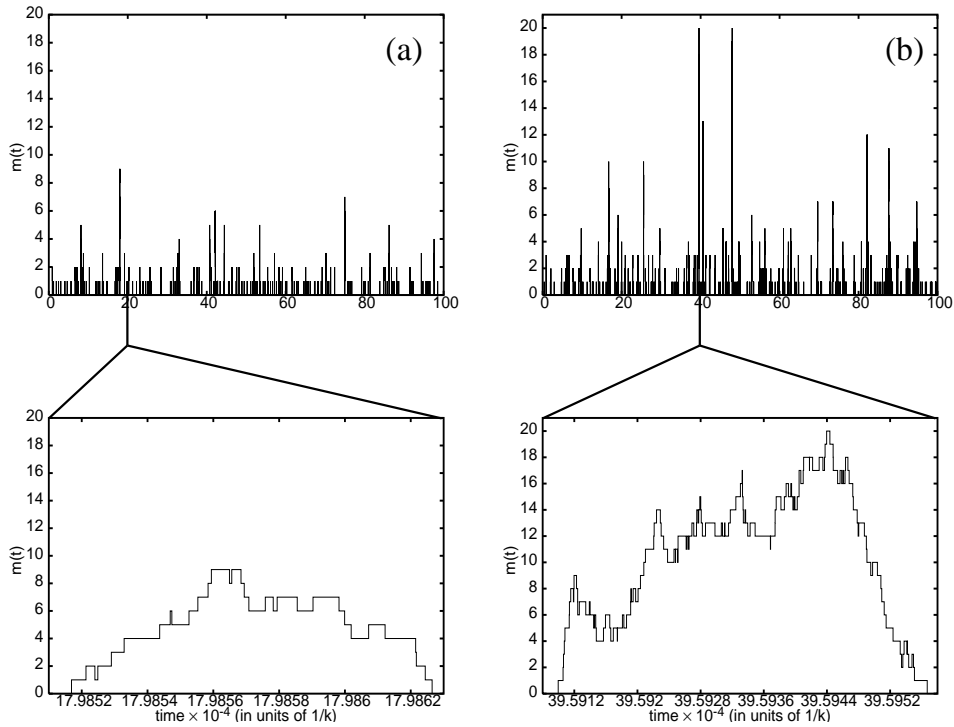}
\caption{Time series of single bubble-breathing dynamics for
$\sigma_0=10^{-3}$, $M=20$, and (a) $u=0.6$ and (b) $u=0.9$. The lower
panel shows a zoom-in of how single bubbles of size $m(t)$ open up and close.}
\label{fig2}
\end{figure}

We start the simulations from the completely zipped state, $b_0=1$ at
$t=0$, and measure the bubble size at time $t$ in terms of
$m(t)=\sum_{m=0}^Mmb_m(t)$. The time series of $m(t)$ for a single stochastic
realization is shown in figure \ref{fig2}. It is distinct that the bubble
events are very sharp (note the time windows of the zoom-ins), and
most of the time the zero-bubble state $b_0$ prevails due to $\sigma_0\ll 1$.
Moreover, raising the temperature increases the bubble size and lifetime, as
it should. By construction of the simulation procedure, it is guaranteed that an
occupation number $b_m=1$ ($m\neq 0$) corresponds to exactly one bubble.

\begin{figure}
\onefigure[scale=0.35,angle=0]{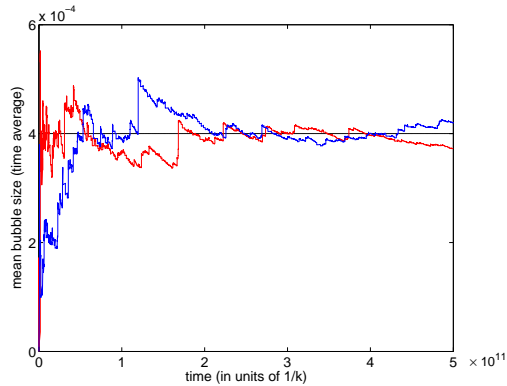}
\caption{Running average of the bubble size $\overline{m}(t)$ for two
realizations (rugged lines). The straight line represents the ensemble
average $\langle m\rangle\approx 4.05\cdot 10^{-4}$ for $\sigma_0=10^{-3}$,
$u=0.6$, and $M=20$.}
\label{fig3}
\end{figure}

To study the ergodicity of the stochastic simulation, we compare the running
average
\begin{equation}
\label{runav}
\overline{m}(t)=t^{-1}\sum_{i=0}^{N-1}\Delta t_im(t_i): \,\,\,\Delta t_{i<
N-1}=t_{i+1}-t_i, \,\, \Delta t_{N-1}=t-t_{N-1}, \,\, t\le
t_N,\,\, t_0=0
\end{equation}
of the bubble size with the ensemble
average, $\langle m\rangle=(\sigma_0/\mathscr{N}_1)\sum_{m=1}^Mmu^m(1+m)^{-c}$,
where $\mathscr{N}_1=1+\sigma_0\sum_{m=1}^Mu^m(1+m)^{-c}$. Note that in
equation (\ref{runav}), we sample over the stochastic time steps $t_i$ chosen
from the reaction PDF (\ref{rpdf}), such that we need to weight the individual
$m(t_i)$ by the time span $\Delta t_i$ until the next jump to $m(t_{i+1})$.
For long times, the quantity (\ref{runav}) is expected to reach the
equilibrium value given by the ensemble-average, i.e., $\overline{m}_{\infty}
\equiv\lim_{t\to\infty}\overline{m}(t)=\langle m\rangle$. In figure
\ref{fig3}, we display the time evolution of $\overline{m}(t)$
for two different stochastic trajectories. Both approach the ensemble-average
$\langle m\rangle$ for longer times. Relatively large deviations from
$\langle m\rangle$ occur, corresponding to a lumping of small or large bubble
states.

\begin{figure}
\twoimages[scale=0.56]{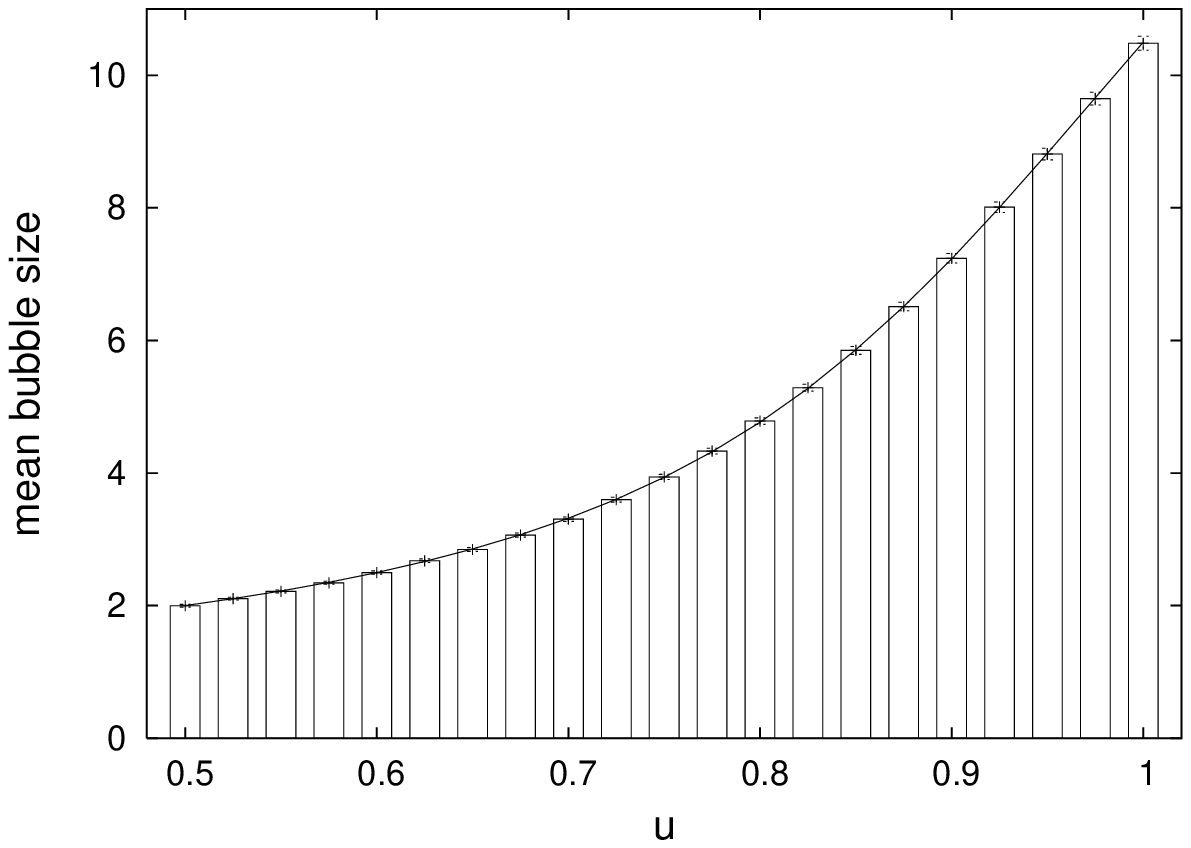}{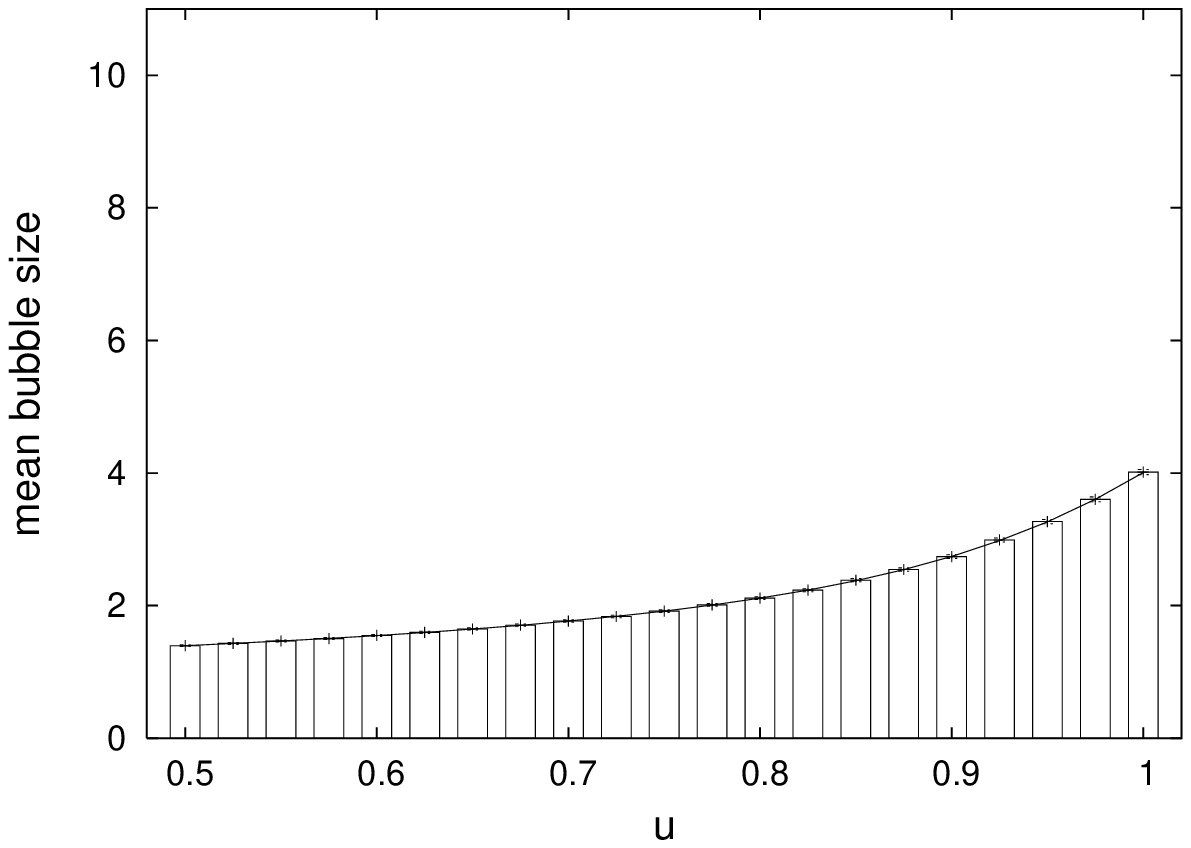}
\caption{Mean bubble size measured exclusively over \emph{open} bubble
states, as a function of statistical weight
$u$ with $c=0$ (left) and $c=1.76$ (right). The boxes represent the time
average from the simulations, the full line is the ensemble-average.
The parameters are $\sigma_0=10^{-3}$ and $M=20$.}
\label{fig4}
\end{figure}

In figure \ref{fig4}, we show the bubble size averaged \emph{exclusively} over
time steps $\Delta t_i$ during which $m(t_i)\neq 0$, as a function of the
statistical weight
$u$. As expected, the bubble size increases with increasing temperature, and
the incorporation of the loop correction term distinctly reduces the maximum
bubble size at equilibrium. The small error bars for the stochastic simulation
data and the good agreement with the theoretical prediction $\sum_{m=1}^Mmu^m
(1+m)^{-c}/\sum_{m=1}^Mu^m(1+m)^{-c}$ demonstrate the reliability of the
Gillespie algorithm.

\begin{figure}
\twoimages[scale=0.56]{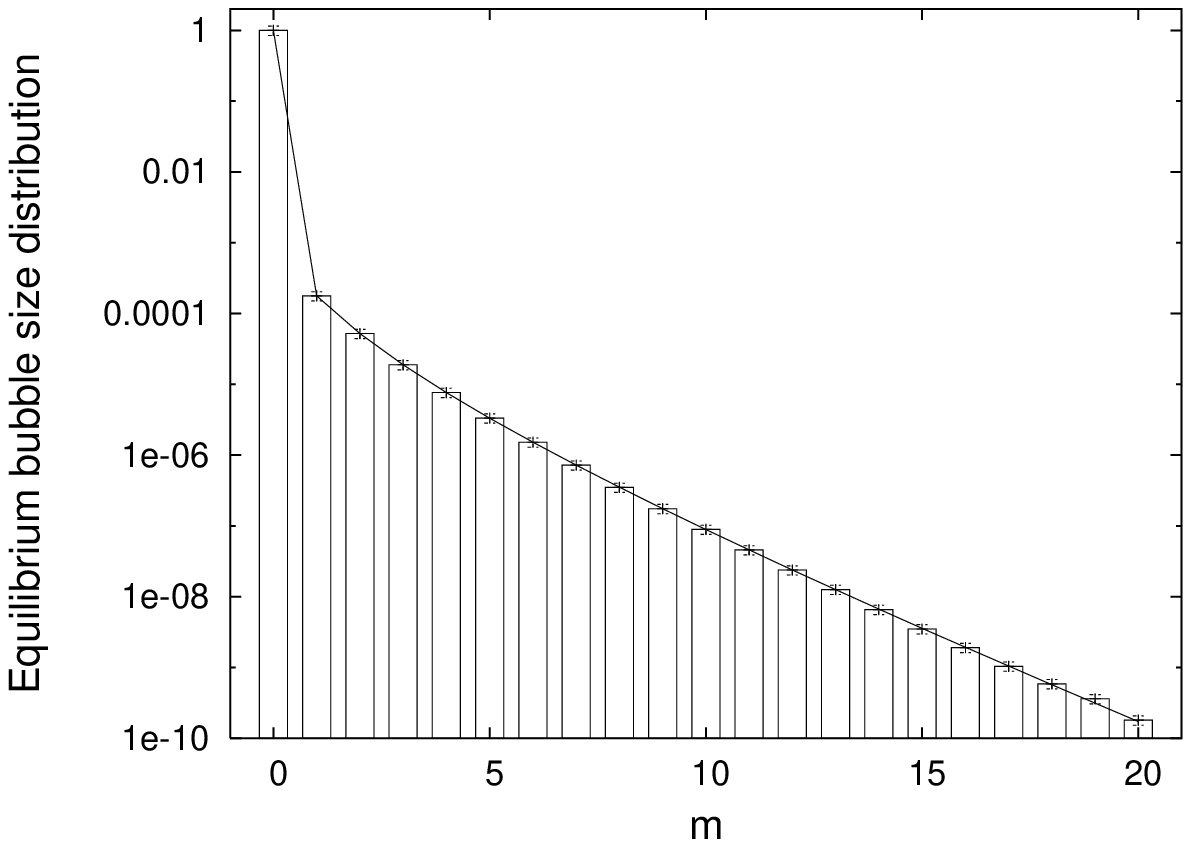}{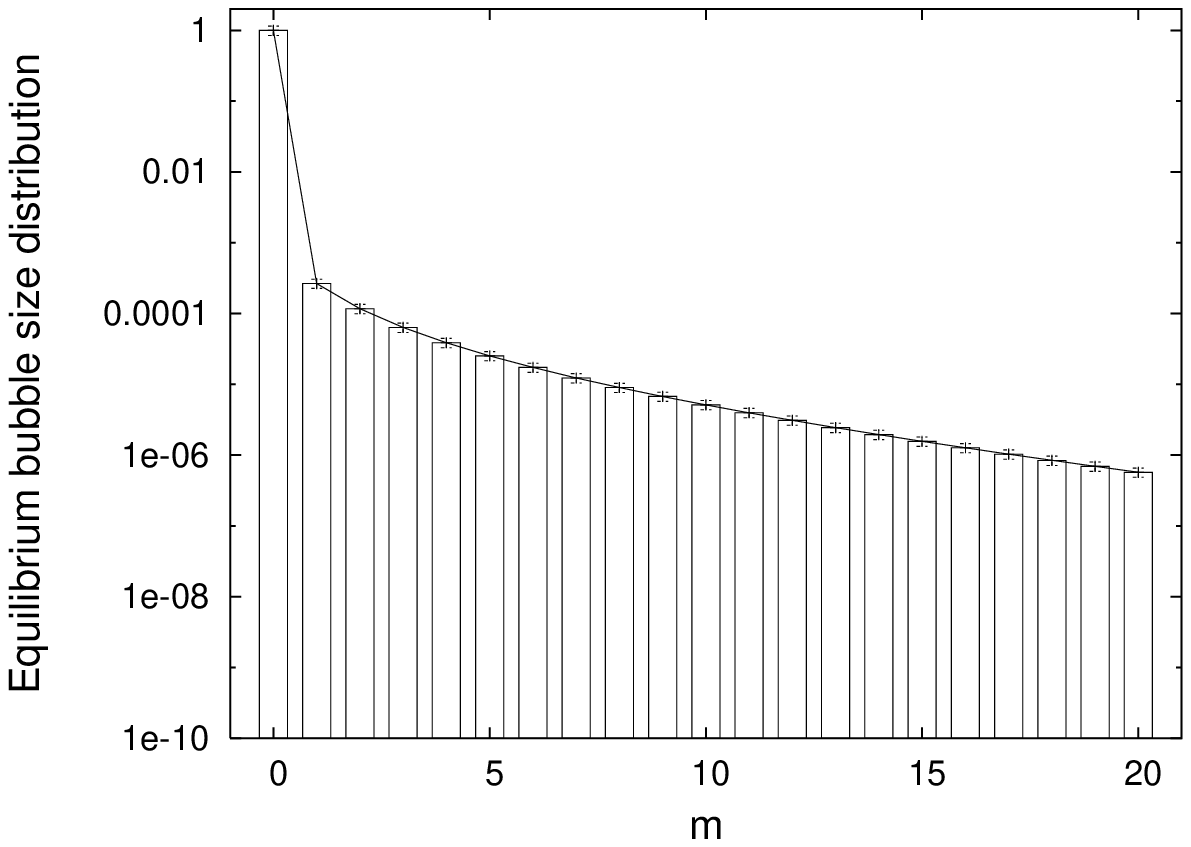}
\caption{Equilibrium bubble size distribution for $u=0.6$ (left) and $u=0.9$
(right) ($\sigma_0=10^{-3}$). The boxes represent the simulations result
($\overline{P}(m,t)$), the full line is the theoretical prediction ($P_{
\mathrm{eq}}(m)$).}
\label{fig5}
\end{figure}

Figure \ref{fig5} depicts the time-averaged bubble size distribution
\begin{equation}
\overline{P}(m,t)=t^{-1}\sum_{i=0}^{N-1}b_m(t_i)\Delta t_i,
\end{equation}
for runs over a large number of jumps $N$ (see below) to ensure that
equilibrium is reached: $\lim_{t\to\infty}\overline{P}(m,t)=P_{\mathrm{eq}}
(m)$. $\overline{P}(m,t)$ is compared to the ensemble-averaged
bubble distribution, $P_{
\mathrm{eq}}(m \geqslant 1)=(1/\mathscr{N}_2)\sigma_0 u^m(1+m)^{-c}$,
with $\mathscr{N}_2=1+\sigma_0\sum_{m=1}^Mu^m(1+m)^{-c}$ and $P_{\mathrm{eq}}
(0)=(1/\mathscr{N}_2)$. This analysis demonstrates that the dsDNA segment
almost always remains completely zipped (since $\sigma_0\ll 1$), note the
logarithmic ordinate. At room temperature the formation of
bigger bubbles is a rare event. Conversely, near the melting temperature
bubble formation is significantly increased (see also figure \ref{fig2}b).
Note that in figures \ref{fig4} and \ref{fig5}, we chose the ordinates such
that they span the same range for the simulated parameter values,
to facilitate easy comparison. Again, small error bars
and good agreement of the time-average with the theoretical ensemble-average
are distinct. The Gillespie algorithm governing our stochastic simulation
therefore reliably leads to relaxation towards equilibrium, as it should, given
that the detailed balance condition is fulfilled by the transfer rate
coefficients $\tr^{\pm}(m)$.

\begin{figure}
\onefigure[scale=0.56,angle=0]{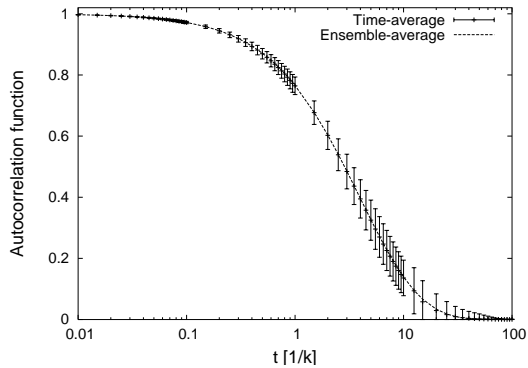}
\caption{Time-averaged ($\overline{C}(t)$) and ensemble-averaged
($C_{\mathrm{eq}}(t)$, see \protect\cite{ambme,ambme1})
autocorrelation functions versus correlation time for
$\sigma_0=10^{-3}$, $u=0.6$, $c=1.76$, and $M=20$.}
\label{fig6}
\end{figure}

We determine the time-averaged autocorrelation function through the
discretized form
\begin{equation}
\label{eq5}
\overline{C}(t)=\left(\frac{1}{N}\sum_{n=0}^Nm(t+n\tau_{\mathrm{bin}})m
(n\tau_{\mathrm{bin}})-\overline{m}_{\infty}^2\right)\Big/
\left(\frac{1}{N}\sum_{n=0}^Nm^2(n\tau_{\mathrm{bin}})-\overline{m}_{\infty}^2
\right),
\end{equation}
Here, $N$ is the number of sample points taken over the trajectory, with
sampling time increments $\tau_{\mathrm{bin}}=\delta/k$. We chose $\delta
\sim 10^{-4}$. Figure \ref{fig6} displays the bubble size autocorrelation
$\overline{C}(t)$ as obtained from the stochastic
simulation in comparison to the ensemble-averaged correlation
$C_{\mathrm{eq}}(t)=\left(\langle m(t)m(0)\rangle-\langle m\rangle^2\right)
/\left(\langle m^2\rangle-\langle m\rangle^2\right)$; the latter is obtained
by numerically solving the eigenvalue provlem associated with the master
equation, see references \cite{ambme,ambme1}).
Both show good agreement. From $\overline{C}(t)$, we conclude that at $u=0.6$,
the typical bubble lifetime should be of the order of $20/k$. Given the typical
experimental bubble lifetime $\simeq 1$ msec at room
temperature, we extract that the zipping rate $k\simeq 50\,\mu$sec, consistent
with reference \cite{altan}.

\section{Implementation}

To determine the random waiting time $\tau$ and the random reaction channel
$\mu$ in the jump process (\ref{eq1}) controlled by the reaction PDF $P(\tau,
\mu)$, two independent random numbers $r_1$ and $r_2$ ($r_i\in[0,1]$)
are generated. In the "direct method", one conditions the reaction PDF in the
form $P(\tau,\mu)=\psi(\tau)P_c(\mu|\tau)$, where $P_c(\mu|\tau)$ is the
probability that the next reaction is a $\mu$ jump, given that the next
reaction occurs at time $t+\tau$ \cite{gillespie}. The waiting time $\tau$
can then be obtained through the relation \cite{gillespie}
\begin{equation}
\tau=\tau(m)=\left(1/\left[\mathsf{t}^+(m)+\mathsf{t}^-(m)\right]\right)
\ln(1/r_1).
\end{equation}
From $r_2$, the direction of the jump is determined as
$\mu=-$, if $0<r_2\left[\mathsf{t}^+(m)+\mathsf{t}^-(m)\right]\le
\mathsf{t}^-(m)$ holds; otherwise, $\mu=+$ \cite{gillespie}.
A typical run would sample over $10^{10}$ to $10^{12}$ jumps to produce
a single time series trajectory, where, in general, lower temperature
requires longer runs. To ensure that equilibrium was reached in such a run,
we checked that the quantity of interest reached (almost) constant values.
For the correlation function, about $10^7$ jumps were sampled at
increments of $10^{-4}/k$. For the time-averages, we calculated the mean
over 100 trajectories.

\section{Conclusions}

By stochastic simulation based on the Gillespie algorithm, we obtained
the time series of the breathing fluctuations of a single homopolymer
DNA bubble, inspired by well-defined DNA-constructs employed in recent
single molecule fluorescence correlation experiments.
The waiting time for a jump event corresponding to a single zipping or
unzipping event defined by the reaction PDF is Poissonian;
however, due to the Boltzmann
factors in the rates, the mean waiting time for a jump to occur can
become quite long.

To corroborate that the stochastic simulation properly describes
thermalization, we determined equilibrium properties such as the mean
bubble size and its distribution from time-averaging, to find almost perfect
agreement with the calculated ensemble-averaged values. We
also showed that the bubble size
autocorrelation function and the running average of the bubble size
follow the behaviour predicted by the master equation. The stochastic
scheme is therefore a reliable way to describe the bubble dynamics.
Computationally, the scheme is quite efficient, and it can be implemented in
a straightforward manner.

The major advantage of the stochastic approach, similar to the Langevin
picture in diffusion processes, is the possibility to sample single
stochastic trajectories and thereby study the direct effect of changes
in the physical parameters (initiation factor $\sigma_0$, statistical weight
$u$, loop closure factor $c$) on the single bubble time series. This is
particularly relevant as it is possible, with the current
experimental means, to record single bubble time series. Given the good
convergence of the statistical sampling, comparison of dynamical data as
obtained from our algorithm to experimental or precise microscopic
simulations data will be an important way to explore further the
validity of the Poland-Scheraga approach to DNA-breathing, in particular,
the exact value of the loop closure exponent $c$ and the bubble initiation
factor $\sigma_0$, as well as potential corrections in the Poland-Scheraga
model for small bubble domains.

Further advantages are the relatively straightforward possibility to
include types of waiting times that are different from the Poissonian, for
instance long-tailed ones \cite{restaurant}. The stochastic simulation is
also more flexible to easily incorporate additional features such as the
coupling to the binding dynamics of selectively single-stranded binding
proteins to the fluctuating bubbles, sequence heterogeneity, or multiple
bubbles, due to the formulation of an arbitrary number of coupled reactions
in the Gillespie algorithm. The corresponding master equation approach in
such cases quickly becomes intractable, once the number of variables
exceeds two. It may therefore be desirable to have available a stochastic
simulation package designed for DNA-breathing similar to the program MELTSIM
for DNA-melting simulations \cite{blake}.

\acknowledgments

We acknowledge very helpful discussions with Andreas Isacsson, and thank
Petter Urkedal for assistance with the Intel Fortran compiler.

\end{document}